\documentclass[a4paper,12pt]{article}
\usepackage{amsthm}
\usepackage{amssymb}
\usepackage{amsmath}
\usepackage{amsfonts}
\usepackage{color}
\usepackage{indentfirst} 
\usepackage{graphicx}

\newtheorem{remark}{Remark}[section]

\newtheorem{lemma}{Lemma}[section]
\newtheorem{theorem}{Theorem}[section]
\newtheorem{proposition}{Proposition}[section]
\newtheorem{corollary}{Corollary}[section]

\def\b1{\mbox{\boldmath $1$}}

\newenvironment{demo*}{\vspace{3mm}\noindent{\bf Proof.}}{\hfill $\Box$ \vspace{3mm}}

\begin{document}
\title{\bf \Large { Stochastic  comparisons of   sample  mean differences for  multivariate   random variables }}
{\author{\normalsize{Xuehua Yin,\,\, Dan Zhu\thanks{Corresponding author: zhudanspring@qfnu.edu.cn},\;Chuancun Yin }\\{\normalsize\it (School of Statistics and Data Science, Qufu Normal University}\\\noindent{\normalsize\it Qufu 273165, Shandong, China)}\\
 \\}
\date{}
\maketitle
\vskip0.01cm
\centerline{\large {\bf Abstract}} The Gini mean difference  is a fundamental statistical measure of dispersion used in many   fields.
 This paper establishes    the usual stochastic orders  and increasing convex orders between the Gini  mean differences  for multivariate
elliptical   random variables which generalized the corresponding results  for multivariate normal  random variables.   We also study the tail probability of the Gini  mean difference  for    multivariate elliptical  random variable  and  revise  a large deviation result for the Gini   mean difference of multivariate normal  random variable in Kim and Kim (2019).

\noindent{\bf Mathematics subject classification (2010)}: 60E15, 62P05.

\noindent{\bf Keywords and phrases:}  {\rm Gini  mean difference; increasing convex order;  large deviation;
 multivariate elliptical   distribution; multivariate normal   distribution; usual stochastic order}


\numberwithin{equation}{section}
\section{Introduction}\label{intro}

Since Corrado Gini introduced an index to
measure concentration or inequality of incomes (see Gini (1936)), it has been   studied extensively because of its importance in various fields such as  economics, actuarial science, finance, operations research, queuing theory and statistics; see, for example,  Denuit et al.   (2005), McNeil et al. (2005),  Brazauskas et al. (2007), Goovaerts et al. (2010), Frees et al. (2011, 2014), Samanthi et al. (2016, 2017),  Kim and Kim (2019),  to name but a few. Recently, there is a growing interest in the study and applications of the  Gini indexes; see e.g., Yin et al. {\color{blue}(2023)}, Yin (2024), Vila et al.  (2024),   Capaldo and  Navarro (2025) and the references therein.

Let $X$ and $Y$ be two independent random variables
on real line with identical distribution $F$, the Gini mean difference is defined
as (see, e.g.,  Gastwirth (1972), Giorgi and Gigliarano (2017))
$$R_G(F)=\frac12E|X-Y|.$$
A commonly used estimator of this index is the sample   mean difference (also called Gini's  mean difference, or Gini index)
{\color{blue} $$G({\bf X})=\frac{1}{2n^2}\sum_{1\le i,j\le n}|X_i-X_j|.$$}
where $X_1, . . . , X_n$  are i.i.d. observations from the population.
Let ${\bf X}=(X_1, . . . , X_n)$   be  a random vector with its components representing risk measures for  $n$ portfolios of risks, these portfolios can be independent or dependent.
Brazauskas et al. (2007) and Samanthi et al. (2017) used this index  to
  check whether or not the $n$ risk measures $X_i$'s are all equal.
 For notational convenience, we denote
 $$G_n({\bf X})=\sum_{1{\color{blue}\le i,j\le n}}|X_i-X_j|.$$
 $G_n({\bf X})$ is the scaled Gini mean difference.
 It is easy to see that $G_n({\bf X})$ can be
rewritten in terms of order statistics
$$G_n({\bf X})=2\sum_{i=1}^n(2i-n-1)X_{(i)},$$
where $X_{(i)}$ denotes the $i$th smallest component of ${X_1, \cdots, X_n}$, i.e. $X_{(1)}\le X_{(2)}\le\cdots\le X_{(n)}$.
Note that the expression (2.2) in Samanthi et al. (2016) lose a negative sign before the summation sign.

 The  comparison of Gini mean differences of multivariate
elliptical risks also shows its own independent interest.
 For example,  for the ordering of Gini mean differences of multivariate normal risks, Samanthi et al. (2016) proposed the following conjecture.

{\bf Conjecture} 1. Let random vector ${\bf X} = (X_1, X_2,\cdots, X_n)'$ follow a multivariate normal distribution  $N_n ({\bf 0},{\bf \Sigma})$.
 Then its Gini index $G_n({\bf X})$ decreases in the sense of usual stochastic order   as the covariance matrix  ${\bf \Sigma}$ increases componentwise with diagonal elements
remaining unchanged.

 Samanthi et al. (2016)  pointed out that proving Conjecture 1 is a challenging
task. They partially completes this task and claim that generalizes the conclusion to elliptical distributions, yet still leaves some open
problems. Recently, Kim and Kim (2019) {\color{blue}has shown} that  this  conjecture  is true when $n = 2$. However, this conjecture is not true when $n \ge  3$. By
using the positive semidefinite ordering of covariance matrices, they obtained the usual stochastic
order of the Gini indexes for multivariate normal risks and generalized to the scale mixture of multivariate
 normal risks. In this paper we generalize the main results in Samanthi et al. (2016) and Kim and Kim (2019) from  multivariate normal risks and scale mixture of multivariate normal risks to  multivariate elliptical risks and scale mixture of multivariate elliptical risks.

 The rest of the paper is organized as follows. In the next section, we  introduce some basic notations and review   definitions and properties of stochastic orders and elliptical distributions. Sections 3 and 4  establish the usual stochastic orders  and increasing convex orders between Gini  mean differences for multivariate
elliptical risks.  We  establish  a large deviation result for the Gini  mean difference of multivariate
normal risks with nonzero mean vector in Section 5. Section 6 provides  concluding remarks of the paper.

\section{Preliminaries}

In this section,   we fix the notation that will be
used in the sequel and we recall some well known results about stochastic orders of random variables and elliptical distributions.
Throughout the paper, we use bold letters to denote vectors
or matrices. For example, ${\bf X}' =(X_1,\cdots, X_n)$ is a row vector and
${\bf \Sigma} = (\sigma_{ij})_{n\times n}$  is an $n\times n$ matrix. In particular, the symbol ${\bf 0}_n$ denotes
the  $n$-dimensional column vector with all entries equal to 0,  ${\bf 1}_{n}$  denotes the $n$-dimensional column
vector with all components equal to  1,  and ${\bf 1}_{n\times n}$ denotes the ${n\times n}$ matrix with all entries equal to 1.  Denote  ${\bf O}_{n\times n}$ as the $n \times n$ matrix having all components equal to 0, and  ${\bf I}_n$ as the $n\times n$ identity matrix. For symmetric matrices $A$ and $B$ of the same size, the notion $A\preceq B$ or $B-A  \succeq {\bf O}$ means that $B-A$ is positive semidefinite.

In order to compare Gini indexes, we recall definitions of some
stochastic orders, see,   Denuit et al.   (2005) and  Shaked and Shanthikumar (2007).
Let $X$ and $Y$ be two random variables, $X$ is said to be smaller than $Y$ in usual stochastic order, denoted as $X \le_{st} Y$, if $P(X>t) \le  P(Y > t)$ for all real numbers $t$.  Random vector ${\bf X}$ is said to be smaller than random vector ${\bf Y}$ in
increasing convex order (written  ${\bf X}\le_{icx} {\bf Y}$), if $E[f({\bf X})] \le E[f({\bf Y})]$
for all increasing convex  functions $f:  {\Bbb{R}}^n \rightarrow {\Bbb R}$ such that the expectations exist.  Similarly, one may defined so called increasing concave order (denoted by  ${\bf X}\le_{icv} {\bf Y}$). Obviously, ${\bf X}\le_{icx} {\bf Y}$ if and only if ${\bf -Y}\le_{icv} {\bf -X}$.
A function $f:  {\Bbb{R}}^n \rightarrow {\Bbb R}$ is said to be supermodular
if for any ${\bf x, y}\in {\Bbb R}^n$ it holds that
$$f({\bf x}) + f({\bf y}) \le f ({\bf x} \wedge {\bf y}) + f ({\bf x} \vee {\bf y}),$$
where the operators $\wedge$ and  $\vee$ denote coordinatewise minimum and
maximum respectively. Supermodular functions are also called quasimonotone  or $L$-superadditive. Note that if $f$ is twice differentiable, then $f$ is supermodular if and only if
 $$\frac{\partial f({\bf x}) }{\partial x_i\partial x_j}\ge 0$$
for all $i\neq j$ and ${\bf x}\in\Bbb{R}^n$. If $-f$ is supermodular, then  $f$ is called submodular.
Random vector ${\bf X}$ is said to be smaller than random vector ${\bf Y}$ in
the supermodular order, denoted as ${\bf X}\le_{sm} {\bf Y}$, if $E[f({\bf X})] \le E[f({\bf Y})]$
for any supermodular function $f$ such that the expectations exist.

We next state some basics about elliptical
distributions. Elliptical distributions have been used widely in insurance, finance and  multicriteria decision theory; see, for example,  Owen and Rabinovitch (1983),  Landsman and Valdez  (2003),  Hamada and Valdez (2008),  Landsman et al. (2018), Sha et al. (2019) and Kim and Kim (2019).
 We follow the notation of Cambanis et al. (1981) and  Fang et al. (1990).
Let ${\bf \Psi}_n$ be a class of functions $\psi: [0,\infty) \rightarrow \Bbb{R}$
such that function $\psi(|\bf t|^2), t\in \Bbb{R}^n$  is an $n$-dimensional
characteristic function. It is clear that
${\bf \Psi}_n\subset {\bf \Psi}_{n-1}\cdots\subset {\bf \Psi}_1.$
Denote by ${\bf \Psi}_{\infty}$  the set of characteristic generators that generate an
$n$-dimensional elliptical distribution for an arbitrary $n\ge 1$. That is
${\bf \Psi}_{\infty}=\cap_{n=1}^{\infty}{\bf \Psi}_{n}.$

An $n \times 1$ random vector ${\bf X} = (X_1, X_2,\cdots, X_n)'$ is said to have an elliptically
symmetric distribution if its characteristic function has the form $e^{i{\bf t}'{\boldsymbol \mu}}\phi({\bf t}'{\bf \Sigma}{\bf t})$ for all ${\bf t}\in \Bbb{R}^n $,
  where  $\phi  \in {\bf \Psi}_n$ is called the characteristic generator satisfying $\phi(0)=1$,
$\boldsymbol{\mu}$ ($n$-dimensional vector) is its location parameter    and  $\bf{\Sigma}$ ($n\times n$ matrix with $\bf{\Sigma}\succeq {\bf O}$) is its dispersion matrix (or scale matrix). The mean vector $E({\bf X})$   (if  exists)
coincides with the location vector and the covariance matrix  Cov$({\bf X})$ (if  exists), being $-2\phi'(0){\bf \Sigma}$.
 We shall write
${\bf{X}}\sim E_n ({\boldsymbol \mu},{\bf \Sigma},\phi)$. In particular, when $\phi(u)=\exp(-u/2)$, we get the multivariate normal distribution, and one  writes
${\bf{X}}\sim N_n ({\boldsymbol \mu},{\bf \Sigma})$. It is well known that
$\bf X$ admits the stochastic representation
\begin{eqnarray*}
{\bf X}={\boldsymbol \mu}+R{\bf A}'{\bf U}^{(n)},
\end{eqnarray*}
where ${\bf A}$  is a square matrix such that ${\bf A}'{\bf A}= {\bf \Sigma}$, ${\bf U}^{(n)}$  is uniformly distributed on the unit sphere ${\cal S}^{n-1}=\{{\bf u}\in \Bbb{R}^n: {\bf u}'{\bf u}=1\} $, random variable $R\geq0$ with $R\sim F$ in $[0,\infty)$ represents the generating variate,   and $F$ is called the generating distribution function, $R$ and  ${\bf U}^{(n)}$ are  independent.
In general,   an elliptically distributed random vector ${\bf{X}}\sim E_n ({\boldsymbol \mu},{\bf \Sigma},\phi)$ does not necessarily
possess a density. However, if density of $X$ exists it must be of the form
\begin{eqnarray*}
f({\bf x})=c_n|{\bf \Sigma}|^{-\frac{1}{2}}g(({\bf x}- {\boldsymbol \mu})^{T}{\bf \Sigma}^{-1}({\bf x-{\boldsymbol \mu} })), \;{\bf x}\in \Bbb{R}^n,
\end{eqnarray*}
for some non-negative function $g$ satisfying the condition
$$\int_0^{\infty}z^{\frac{n}{2}-1} g(z)dz<\infty, $$
and a normalizing constant $c_n$ given by
$$c_n=\frac{\Gamma(\frac{n}{2})}{\pi^{\frac{n}{2}}}\left(\int_0^{\infty}z^{\frac{n}{2}-1} g(z)dz\right)^{-1}.$$
  The function $g$ is called the density generator. Sometimes, we
write $X \sim E_n ({\boldsymbol \mu},{\bf \Sigma},g)$ for the $n$-dimensional elliptical distributions generated from the
function $g$. In this case $R$ in (2.1) has the pdf given by
\begin{eqnarray*}
h_R(v)=c_n \frac{2\pi^{\frac{n}{2}}}{\Gamma(\frac{n}{2})}v^{n-1}g(v^2), v\ge 0.
\end{eqnarray*}

Theorem 2.21 in Fang et al. (1990) showed that  $\psi\in {\bf \Psi}_{\infty}$ if and only if  ${\bf{X}}\sim {E}_n({\boldsymbol \mu},{\bf \Sigma},{\bf \psi})$   is a mixture of normal distributions. Some such elliptical distributions are the multivariate normal distribution, the multivariate $T$-distribution, the multivariate Cauchy distribution and  the exponential power distribution $EP_n({\boldsymbol \mu},{\bf \Sigma},\beta)$  with $\beta\in (0,1]$. Some elliptical distributions like logistic distribution and Kotz type distribution are not mixture of normal distributions.
A comprehensive review of the properties and characterizations of elliptical and related distributions can be found  in  Cambanis et al. (1981),  Fang et al. (1990) and  Adcock and Azzalini (2020).

\section{Usual stochastic order of Gini indexes}

In this section,  we extend  the results of multivariate
normal risks and scale mixture  multivariate
normal risks in  Samanthi et al. (2016) and  Kim and Kim (2019)  to scale mixture multivariate elliptical
risks. To compare the usual
stochastic orders between Gini indexes for multivariate
elliptical risks, we
use the following result due to  Fefferman et al. (1972); see   Eaton and Erlman (1991) for a different proof. In the case of  Gaussian distribution,
  the corresponding result is proved   by Anderson (1955).
Let ${\cal{C}}$ denote the class of all convex, centrally symmetric (i.e., $C =-C$)
subsets $C$ of ${\Bbb{R}^n}$.

\begin{lemma} Suppose that ${\bf{X}}\sim E_n ({\bf 0},{\bf \Sigma}_{x},\phi)$ and ${\bf{Y}}\sim E_n ({\bf 0},{\bf \Sigma}_{y},\phi)$. If ${\bf \Sigma}_{x}  \preceq {\bf \Sigma}_{y}$, then  for every $C \in {\cal C}$, $$P({\bf X}\in C)\ge  P({\bf Y}\in C).$$
\end{lemma}
The following result generalized Proposition 2 in Kim and Kim (2019) in which they only {\color{blue}considered}  a special class of multivariate elliptical risks with zero mean vector, i.e., scale mixture of multivariate  normal risks with zero mean vector.
\begin{proposition}
Let ${\bf{X}}\sim E_n ( {\boldsymbol \mu},{\bf \Sigma}_{x},\phi)$ and ${\bf{Y}}\sim E_n ({\boldsymbol \mu},{\bf \Sigma}_{y},\phi)$. If ${\bf \Sigma}_{x}  \preceq {\bf \Sigma}_{y}$, then
\begin{eqnarray*}
   G_n({\bf X}-{\boldsymbol \mu})\le_{st} G_n({\bf Y}-{\boldsymbol \mu}).
\end{eqnarray*}
In particular, if ${\boldsymbol \mu}={\bf 0}$, or $\mu{\bf 1}_n$, then
 \begin{eqnarray*}
 G_n({\bf X})\le_{st} G_n({\bf Y}).
 \end{eqnarray*}
\end{proposition}
{\bf Proof.}\; If ${\bf{X}}\sim E_n ( {\boldsymbol \mu},{\bf \Sigma}_{x},\phi)$ and ${\bf{Y}}\sim E_n ({\boldsymbol \mu},{\bf \Sigma}_{y},\phi)$, then ${\bf{X}}-{\boldsymbol \mu}\sim E_n ( {\bf 0},{\bf \Sigma}_{x},\phi)$ and ${\bf{Y}}-{\boldsymbol \mu}\sim E_n ({\bf 0},{\bf \Sigma}_{y},\phi)$, and thus (3.1) follows from Lemma 3.1 by taking $C_t=\{{\bf x}\in {\Bbb{R}}^n: G_n({\bf x})\le t\}$ for $t>0$ as in Kim and Kim (2019).
It is easy to check that the Gini index $G_n(\cdot)$ is invariant under  drift  $\mu{\bf 1}_n$, i.e.,  $G_n({\bf X}+ \mu{\bf 1}_n)=G_n({\bf X})$ for all $n$-dimensional random vector ${\bf X}$.  Therefore,   (3.2)  follows. $\hfill\square$

We will   extend the result of Proposition 3.1 to  the {\color{blue}scale}   mixture of multivariate elliptical risks.

{\bf Definition 2.1} A $n$-dimensional random variable ${\bf X}$ is said to have a  scale mixture of elliptical distributions with the parameters ${\boldsymbol \mu}$ and  ${\bf \Sigma}$, if
\begin{eqnarray*}
{\bf X}={\boldsymbol \mu}+\sqrt{V}{\bf \Sigma}^{\frac12}{\bf Z},
\end{eqnarray*}
where ${\bf Z} \sim ELL_n ({\bf 0},{\bf I}_n,\phi)$, $V$ is a nonnegative, scalar-valued random variable with the distribution $F$, ${\bf Z}$ and $V$ are independent,    ${\boldsymbol \mu}\in \Bbb{R}^n$,  ${\bf \Sigma}\in\Bbb{R}^{n\times n}$ with ${\bf \Sigma}\succeq{\bf O}$, and  ${\bf \Sigma}^{\frac12}$ is the square root of  ${\bf \Sigma}$. Here ${\bf 0}$ is an $n\times1$ vector of zeros.  We will use the notation ${\bf Y}\sim SME_n({\boldsymbol \mu},{\bf \Sigma},\phi;F)$.

Note that  when ${\bf Z} \sim N_n(0,{\bf I_n})$ we get the multivariate normal variance mixture distribution (see, e.g., McNeil et al.,
2005);
When ${\bf Z} \sim KTD_n(0,{\bf I_n}, N,\frac12, \beta)$ we have  the variance mixture of the Kotz-type distribution  introduced by Arslan (2009).

 Proposition 3.1 can be generalized to the  scale mixture of multivariate elliptical risks.
\begin{proposition}
 Let ${\bf{X}}\sim SME_n({\boldsymbol \mu},{\bf \Sigma}_x,\phi;F)$
and ${\bf{Y}}\sim SME_n({\boldsymbol \mu},{\bf \Sigma}_y,\phi;F)$. If ${\bf \Sigma}_{x}  \preceq {\bf \Sigma}_{y}$, then
\begin{eqnarray*}
   G_n({\bf X}-{\boldsymbol \mu})\le_{st} G_n({\bf Y}-{\boldsymbol \mu}).
   \end{eqnarray*}
   In particular, if  ${\boldsymbol \mu}={\bf 0}$, or $\mu{\bf 1}_n$, then
 \begin{eqnarray*}
 G_n({\bf X})\le_{st} G_n({\bf Y}).
 \end{eqnarray*}
\end{proposition}
{\bf Proof.}\; It can be easily seen that  for any $v>0$,   ${\bf X}|V=v \sim E_n({\boldsymbol \mu},v{\bf \Sigma}_x,\phi)$ and  ${\bf Y}|V=v\sim E_n({\boldsymbol \mu},v{\bf \Sigma}_y,\phi)$. Since ${\bf \Sigma}_{x}  \preceq {\bf \Sigma}_{y}$, one has
$v{\bf \Sigma}_{x}  \preceq v{\bf \Sigma}_{y}$. By Proposition 3.1, given $V=v$, we get
$$G_n({\bf X}-{\boldsymbol \mu})\le_{st} G_n({\bf Y}-{\boldsymbol \mu}).$$
Or, equivalently,
$$G_n(\sqrt{v}{\bf \Sigma}_x^{\frac12}{\bf Z})\le_{st} G_n( \sqrt{v}{\bf \Sigma}_y^{\frac12}{\bf Z}).$$
Therefore, for all $t\in {\Bbb{R}}$,
\begin{eqnarray*}
P(G_n({\bf X}-{\boldsymbol \mu})>t)&=&P(G_n(\sqrt{V}{\bf \Sigma}_x^{\frac12}{\bf Z} )>t)\\
&=&\int_0^{\infty}P(G_n(\sqrt{v}{\bf \Sigma}_x^{\frac12}{\bf Z} )>t)dF(v)\\
&\le& \int_0^{\infty}P(G_n(\sqrt{v}{\bf \Sigma}_y^{\frac12}{\bf Z} )>t)dF(v)\\
&=&P(G_n(\sqrt{V}{\bf \Sigma}_y^{\frac12}{\bf Z} )>t)\\
&=&P(G_n({\bf Y}-{\boldsymbol \mu})>t),
\end{eqnarray*}
which is (3.4).
In particular, if  ${\boldsymbol \mu}={\bf 0}$, or $\mu{\bf 1}_n$, then $G_n({\bf X}-{\boldsymbol \mu})=G_n({\bf X})$ and
$G_n({\bf Y}-{\boldsymbol \mu})=G_n({\bf Y})$, and (3.5) follows. $\hfill\square$

An important property of elliptical distributions is that linear transformations of elliptical
vectors are also ellipticals with the same characteristic generator. Specifically, we conclude this property in the following lemma.
\begin{lemma} (Fang et al. (1990)) Suppose that ${\bf{X}}\sim E_n ( {\boldsymbol \mu},{\bf \Sigma},\phi)$, ${\bf B}$  is an $m\times n$ matrix  of rank $m\le n$, and ${\bf b}$ is an $m\times 1$ vector, then ${\bf BX+b}\sim E_m ({\bf B}{\boldsymbol \mu}+{\bf b},{\bf B}{\bf \Sigma}{\bf B}',\phi)$.
\end{lemma}

We will give a weaker  sufficient condition  for stochastic ordering of Gini indexes for multivariate elliptical risks.

\begin{proposition}
Let ${\bf{X}}\sim E_n ( {\boldsymbol \mu},{\bf \Sigma}_{x},\phi)$ and ${\bf{Y}}\sim E_n ({\boldsymbol \mu},{\bf \Sigma}_{y},\phi)$. If ${\bf A}{\bf \Sigma}_{x}{\bf A}'  \preceq {\bf A} {\bf \Sigma}_{y}{\bf A}'$,
where ${\bf A}$ is
 an $n\times n$ matrix defined as
$${\bf A}=\left(\begin{array}{cccc}
 \frac{n-1}{n}\;& -\frac{1}{n}\; &\cdots\; &-\frac{1}{n} \\
 -\frac{1}{n}\; &\frac{n-1}{n}\; &\cdots\;  &-\frac{1}{n}\\
 \vdots\;&\vdots\;&\ddots\;&\vdots\;\\
 -\frac{1}{n}\; &-\frac{1}{n}\; &\cdots\;  &\frac{n-1}{n}\\
\end{array}\right).
$$
Then
\begin{eqnarray*}
   G_n({\bf AX}-{\bf A \boldsymbol  \mu})\le_{st}    G_n({\bf AY}-{\bf A \boldsymbol  \mu}).
\end{eqnarray*}
In particular, if ${\boldsymbol \mu}={\bf 0}$,   or $\mu{\bf 1}_n$, then
 \begin{eqnarray*}
 G_n({\bf X})\le_{st} G_n({\bf Y}).
 \end{eqnarray*}
\end{proposition}

 {\bf Proof.}\; If ${\bf{X}}\sim E_n ( {\boldsymbol \mu},{\bf \Sigma}_{x},\phi)$ and ${\bf{Y}}\sim E_n ({\boldsymbol \mu},{\bf \Sigma}_{y},\phi)$,
 by Lemma 3.2 we get
 $${\bf{AX}}\sim E_n ( {\bf A}{\boldsymbol \mu}, {\bf A}{\bf \Sigma}_{x}{\bf A}',\phi)$$
  and
  $${\bf{AY}}\sim E_n ({\bf A}{\boldsymbol \mu},{\bf A}{\bf \Sigma}_{y}{\bf A}',\phi).$$
According to Proposition 3.1, if  ${\bf A}{\bf \Sigma}_{x}{\bf A}'  \preceq {\bf A} {\bf \Sigma}_{y}{\bf A}'$, then
$$ G_n({\bf AX}-{\bf A \boldsymbol  \mu})\le_{st}    G_n({\bf AY}-{\bf A \boldsymbol  \mu}).$$
In particular, if ${\boldsymbol \mu}={\bf 0}$, or $\mu{\bf 1}_n$, then ${\bf A \boldsymbol  \mu}={\bf 0}$. Thus
$$ G_n({\bf AX})\le_{st}    G_n({\bf AY}),$$
and (3.7) follows since $G_n({\bf AX})= G_n({\bf X})$ and $G_n({\bf AY})= G_n({\bf Y})$. $\hfill\square$

The following proposition  generalizes the result of  Proposition 4.4 in Samanthi et al. (2016) in which only  multivariate normal risks with zero mean vectors are considered. Moreover, we provide a short proof.
\begin{proposition}
Let ${\bf{X}}\sim E_n ({\boldsymbol \mu},{\bf \Sigma}_{x},\phi)$ and ${\bf{Y}}\sim E_n ({\boldsymbol \mu},{\bf \Sigma}_{y},\phi)$. If
there exists $\varepsilon\in {\Bbb R}$ such that ${\bf \Sigma}_{y}-{\bf \Sigma}_{x}+\varepsilon {\bf 1}_{n\times n}\succeq {\bf O}_{n\times n}$,
then
\begin{eqnarray*}
   G_n({\bf AX}-{\bf A \boldsymbol  \mu})\le_{st}    G_n({\bf AY}-{\bf A \boldsymbol  \mu}),
\end{eqnarray*}
where ${\bf A}$ is defined in  Proposition 3.3.
In particular, if ${\boldsymbol \mu}={\bf 0}$,   or $\mu{\bf 1}_n$, then
 \begin{eqnarray*}
 G_n({\bf X})\le_{st} G_n({\bf Y}).
 \end{eqnarray*}
\end{proposition}
{\bf Proof.}\; By Proposition 3.3, it suffices to show that  ${\bf A}{\bf \Sigma}_{x}{\bf A}'  \preceq {\bf A} {\bf \Sigma}_{y}{\bf A}'$.
In fact, $${\bf A} {\bf \Sigma}_{y}{\bf A}'-{\bf A}{\bf \Sigma}_{x}{\bf A}'={\bf A} ({\bf \Sigma}_{y}-{\bf \Sigma}_{x} +\varepsilon {\bf 1}_{n\times n}){\bf A}'\succeq  {\bf O}_{n\times n},$$
since ${\bf A} {\bf 1}_{n\times n}{\bf A}'={\bf 0}$. $\hfill\square$

\begin{remark} If ${\bf \Sigma}_{x}\preceq {\bf \Sigma}_{y}$ for any $\varepsilon\ge 0$, we have
${\bf \Sigma}_{y}-{\bf \Sigma}_{x}+\varepsilon {\bf 1}_{n\times n}\succeq  {\bf O}_{n\times n}$. But conversely   is not true in general.
\end{remark}

Proposition 3.5 can be generalized to   scale mixture of  multivariate elliptical risks. The proof is similar to that used in extending
Propositions 3.1 to 3.2 and hence is omitted.

\begin{proposition}
 Let ${\bf{X}}\sim SME_n({\boldsymbol \mu},{\bf \Sigma}_x,\phi;F)$
and ${\bf{Y}}\sim  SME_n({\boldsymbol \mu},{\bf \Sigma}_y,\phi;F)$. If ${\bf A}{\bf \Sigma}_{x}{\bf A}'  \preceq {\bf A} {\bf \Sigma}_{y}{\bf A}'$,
where ${\bf A}$ is defined in  Proposition 3.3, then
  \begin{eqnarray*}
   G_n({\bf AX}-{\bf A \boldsymbol  \mu})\le_{st}    G_n({\bf AY}-{\bf A \boldsymbol  \mu}).
\end{eqnarray*}
In particular, if ${\boldsymbol \mu}={\bf 0}$,  or $\mu{\bf 1}_n$, then
 \begin{eqnarray*}
 G_n({\bf X})\le_{st} G_n({\bf Y}).
 \end{eqnarray*}
\end{proposition}

\begin{proposition}
 Let ${\bf{X}}\sim SME_n({\boldsymbol \mu},{\bf \Sigma}_x,\phi;F)$
and ${\bf{Y}}\sim  SME_n({\boldsymbol \mu},{\bf \Sigma}_y,\phi;F)$.
If there exists $\varepsilon\in {\Bbb R}$ such that ${\bf \Sigma}_{y}-{\bf \Sigma}_{x}+\varepsilon {\bf 1}_{n\times n} \succeq  {\bf O}_{n\times n}$,
 then
  \begin{eqnarray*}
   G_n({\bf AX}-{\bf A \boldsymbol  \mu})\le_{st}    G_n({\bf AY}-{\bf A \boldsymbol  \mu}).
\end{eqnarray*}
In particular, if ${\boldsymbol \mu}={\bf 0}$,  or $\mu{\bf 1}_n$, then
 \begin{eqnarray*}
 G_n({\bf X})\le_{st} G_n({\bf Y}),
 \end{eqnarray*}
 where ${\bf A}$ is defined in  Proposition 3.3.
\end{proposition}
{\bf Proof.}\; It is an immediate consequence of Proposition 3.5 since the condition ${\bf \Sigma}_{y}-{\bf \Sigma}_{x}+\varepsilon {\bf 1}_{n\times n}\succeq  {\bf O}_{n\times n}$ implies  ${\bf A}{\bf \Sigma}_{x}{\bf A}'  \preceq {\bf A} {\bf \Sigma}_{y}{\bf A}'$ as shown in the proof of Proposition 3.4. $\hfill\square$

The condition on the components of dispersion matrix of   multivariate normal risk or scale   mixture of multivariate normal risk ${\bf X}$ for  the monotonicity of the Gini index $G_n({\bf X})$ in the usual stochastic order   proposed by  Kim and Kim (2019) is also  suitable for the general    multivariate elliptical risk or scale   mixture of multivariate elliptical risk, as shown below.

The following result generalizes Propositions 3 and 4 in Kim and Kim (2019) in which they only considered  a special class of multivariate elliptical risks with zero mean vectors, i.e.,  the  multivariate  normal risks and scale mixture of multivariate  normal risks with zero mean vectors.

\begin{proposition}
 Let ${\bf{X}}\sim SME_n({\boldsymbol \mu},{\bf \Sigma}_x,\phi;F)$
and ${\bf{Y}}\sim SME_n({\boldsymbol \mu},{\bf \Sigma}_y,\phi;F)$ with  ${\bf \Sigma}_x=(\sigma^x_{ij})_{n\times n}$ and  ${\bf \Sigma}_y=(\sigma^y_{ij})_{n\times n}$. Let $\varepsilon>0$, if  $\sigma^y_{ij}=\sigma^x_{1j}+\varepsilon, j=2,\cdots,n$,
$\sigma^y_{i1}=\sigma^x_{i1}+\varepsilon, i=2,\cdots,n$ and  for other $1\le i,j\le n$,  $\sigma^y_{ij}=\sigma^x_{ij}$.  Then
\begin{eqnarray*}
   G_n({\bf X}-{\boldsymbol \mu})\ge_{st} G_n({\bf Y}-{\boldsymbol \mu}).
   \end{eqnarray*}
   In particular, if  ${\boldsymbol \mu}={\bf 0}$, or $\mu{\bf 1}_n$, then
 \begin{eqnarray*}
 G_n({\bf X})\ge_{st} G_n({\bf Y}).
 \end{eqnarray*}
\end{proposition}
{\bf Proof.}\; According to  Kim and Kim (2019), under the assumed condition, we know that,
$${\bf \Sigma}_{x}-{\bf \Sigma}_{y}=\varepsilon\left(\begin{array}{cc}
 0\; &-{\bf 1}'_{n-1}\\
 -{\bf 1}_{n-1}\; &{\bf O}_{(n-1)\times (n-1)}\\
\end{array}\right),$$
from which we get
 $${\bf \Sigma}_{x}-{\bf \Sigma}_{y}+\varepsilon {\bf 1}_{n\times n}=\varepsilon\left(\begin{array}{cc}
 1\; &{\bf 0}'_{n-1}\\
 {\bf 0}_{n-1}\; &{\bf 1}_{(n-1)\times (n-1)}\\
\end{array}\right).$$
We conclude that the latter matrix is positive semidefinite. In fact, for any ${\bf x}=(x_1,\cdots,x_n)'\in \Bbb{R}^n$,
$${\bf x}'\left(\begin{array}{cc}
 1\; &{\bf 0}'_{n-1}\\
 {\bf 0}_{n-1}\; &{\bf 1}_{(n-1)\times (n-1)}\\
\end{array}\right){\bf x}=x_1^2+(x_2+\cdots+x_n)^2\ge 0.$$
Therefore, ${\bf \Sigma}_{x}-{\bf \Sigma}_{y}+\varepsilon {\bf 1}_{n\times n}\succeq {\bf O}_{n\times n}.$
Then Proposition 3.6  implies the desired results. $\hfill\square$

The following result generalizes Propositions 5 and 6 in Kim and Kim (2019) in which they only considered    the  multivariate  normal risks and scale mixture of multivariate  normal risks with zero mean vectors.
\begin{proposition}
 Let ${\bf{X}}\sim SME_n({\boldsymbol \mu},{\bf \Sigma}_x,\phi;F)$
and ${\bf{Y}}\sim SME_n({\boldsymbol \mu},{\bf \Sigma}_y,\phi;F)$ with
 ${\bf \Sigma}_x=(\sigma^x_{ij})_{n\times n}$ and  ${\bf \Sigma}_y=(\sigma^y_{ij})_{n\times n}$. Let $\varepsilon>0$,
  if
  \begin{eqnarray*}
 \sigma^y_{ij}=\left\{\begin{array}{ll}  \sigma^x_{ij}+\varepsilon,  \ & {\rm if}\; i\neq j,\\
 \sigma^x_{ij},\ &{\rm if}\;   i=j,
 \end{array}
  \right.
\end{eqnarray*}
  then
\begin{eqnarray*}
   G_n({\bf X}-{\boldsymbol \mu})\ge_{st} G_n({\bf Y}-{\boldsymbol \mu}).
   \end{eqnarray*}
   In particular, if  ${\boldsymbol \mu}={\bf 0}$, or $\mu{\bf 1}_n$, then
 \begin{eqnarray*}
 G_n({\bf X})\ge_{st} G_n({\bf Y}).
 \end{eqnarray*}
\end{proposition}
{\bf Proof.}\;  Under the assumed condition, Kim and Kim (2019) find  that,
$${\bf \Sigma}_{x}-{\bf \Sigma}_{y}=\varepsilon({\bf I}_{n\times n}-{\bf 1}_n{\bf 1}_n'),$$
from which we get
  \begin{eqnarray*}
 {\bf A}({\bf \Sigma}_{x}-{\bf \Sigma}_{y}){\bf A}'&=&\varepsilon {\bf A}{\bf I}_{n\times n}{\bf A}'-\varepsilon{\bf A}{\bf 1}_{n}{\bf 1}'_{n}{\bf A}'\\
 &=& \varepsilon {\bf A}{\bf A}'-\varepsilon{\bf A}{\bf 1}_{n\times n}{\bf A}'\\
 &=&\varepsilon {\bf A}{\bf A}'\succeq {\bf O}_{n\times n},
 \end{eqnarray*}
 where ${\bf A}$ is defined in  Proposition 3.3.
 We find that ${\bf A}{\bf \Sigma}_{x}{\bf A}' \succeq {\bf A}{\bf \Sigma}_{y}{\bf A}'$.
Therefore,  Proposition 3.5    provides the desired result. $\hfill\square$

 \section{Increasing convex order of  Gini indexes}

 Samanthi et al.  (2016) {\color{blue} established } a sufficient and necessary condition
for the supermodular order between two   scale mixture of multivariate normal risks.  Based on this result they
find  a sufficient condition
for the  increasing convex order  of  Gini indexes for  two 3-dimensional  elliptical random variables.  In addition, they remark that ordering  $G_n({\bf X})$
in the increasing convex order for higher dimensional risk is still an open problem.     Since $-G_n({\bf X})$ is  supermodular, but not  componentwise  monotone in general, so that  for any  convex and increasing function  $\psi:  {\Bbb{R}} \rightarrow {\Bbb R}$,  the composition
$\psi(-G_n({\bf X}))$ is not   supermodular in general, a counterexample for 4-dimensional case can be found in  Samanthi et al.  (2016).
Therefore, the above open problem is not true in general. We now consider the supermodular order of  Gini indexes for general elliptical random variables. We first cite Theorem 3.4 of Yin (2019) below, which is an extension of Proposition 3.4 in  Samanthi et al.  (2016).
\begin{proposition}
 Let  ${\bf{X}}\sim E_n ({\boldsymbol \mu}^x,{\bf \Sigma}^x,\phi)$ and   ${\bf{Y}}\sim E_n ({\boldsymbol \mu}^y,{\bf \Sigma}^y,\phi)$ with ${\bf\Sigma}^x=(\sigma^x_{ij})_{n\times n}$ and  ${\bf \Sigma}^y=(\sigma^y_{ij})_{n\times n}$. Then
 ${\bf X}\le_{sm} {\bf Y}$ if and only if ${\bf X}$ and ${\bf Y}$ have the same  marginal  and  $\sigma^x_{ij}\le  \sigma^y_{ij}$ for all $1\le i<j\le n$.
\end{proposition}
The following result is a direct consequence of  Proposition  4.1,  which  generalizes Proposition 3.6  in  Samanthi et al.  (2016).
 \begin{corollary}
 Let ${\bf{X}}\sim E_n({\boldsymbol \mu}^x,{\bf \Sigma}^x,\phi)$
and ${\bf{Y}}\sim E_n({\boldsymbol \mu}^y,{\bf \Sigma}^y,\phi)$ with
 ${\bf \Sigma}^x=(\sigma^x_{ij})_{n\times n}$ and  ${\bf \Sigma}^y=(\sigma^y_{ij})_{n\times n}$.
 If  ${\bf X}$ and ${\bf Y}$ have the same marginals and  $\sigma^x_{ij}\le  \sigma^y_{ij}$ for all $1\le i<j\le n$, then $E(G_n({\bf X}))\ge E(G_n({\bf Y}))$ given the expectations exist.
\end{corollary}

\section{Tail asymptotic results for Gini indexes}
In this section, we will discuss the  asymptotic result for the    tail
probability of Gini index $G_n({\bf X})$ when  ${\bf{X}}\sim {E}_n({\boldsymbol 0},{\bf \Sigma},{\bf \psi})$.
The following lemma can be found in  Fang and Liang (1989), see also Tong (1990).
\begin{lemma} For all real vectors ${\bf X}=(X_1, \cdots, X_n)'$ and all given real numbers $0\le C_1\le\cdots\le C_n$, we have
$$\sum_{i=1}^n C_i X_{(i)}=\sup_{{\bf C_r}\in\Pi}{\bf C_r}'{\bf X},$$
where $X_{(1)}\le X_{(2)}\le\cdots\le X_{(n)}$  is the rearranged order of {\color{blue}$X_1, \cdots, X_n$} and  $\Pi$ is  the $n!$  vectors of    permutations of $( C_1,\cdots, C_n)'$.
\end{lemma}


To discuss the  asymptotic result for the    tail
probability of Gini index $G_n({\bf X})$, we first recall  some results for elliptical distributions.
Consider the linear transformation ${\bf Y}={\bf CX}$, where
$${\bf C}'=({\bf C}_1,\cdots, {\bf C}_m),\; {\bf C}_r=(C_{r_1},\cdots, C_{r_n})'$$
where $m=n!$ and ${\bf C}$ is an $m\times n$ matrix such that ${\bf C_r}$ is a permutation of $(C_{1},\cdots, C_{n})'$.
If  ${\bf{X}}\sim E_n ({\boldsymbol \mu},{\bf \Sigma},\phi)$, then by Lemma 3.2,  ${\bf{Y}}=(Y_1,\cdots,Y_m)' \sim E_m ({\bf C}{\boldsymbol \mu}, {\bf C}{\bf \Sigma}{\bf C}',\phi)$. This together with Lemma 5.1 imply that
$$P\left(\sum_{i=1}^n C_iX_{(i)}\le x\right)=P\left(\cap_{j=1}^m\{Y_j\le x\}\right).$$
This result in the multinormal case can be found in Tong (1990).
In particular, taking $C_i=4i-2n-2, i=1,2,\cdots, n$ leads to the following result.
\begin{proposition}
Assume that  ${\bf{X}}\sim E_n ({\boldsymbol \mu},{\bf \Sigma},\phi)$,   then for any $x>0$,
 \begin{eqnarray*}
P(G_n({\bf X})\le x)=P\left(\cap_{j=1}^m\{Y_j\le x\}\right).
  \end{eqnarray*}
Or, equivalently,
\begin{eqnarray*}
P(G_n({\bf X})>x)=P\left(\cup_{j=1}^m\{Y_j>x\}\right),
  \end{eqnarray*}
where
 ${\bf{Y}}=(Y_1,\cdots,Y_m)' \sim E_m ({\bf C}{\boldsymbol \mu}, {\bf C}{\bf \Sigma}{\bf C}',\phi)$
 with $m=n!$,  $C_i=4i-2n-2, i=1,2,\cdots, n$ and
$${\bf C}'=({\bf C}_1,\cdots, {\bf C}_m).$$
Here ${\bf C}_r$ is a permutation of $(C_{1},\cdots, C_{n})'$, $r=1,2,\cdots, m$.
\end{proposition}
The next theorem improves the result of Theorem 5 in Kim and Kim (2019).
\begin{theorem} Let ${\bf{X}}\sim N_n({\boldsymbol \mu},{\bf \Sigma})$, then we have
$$\lim_{x\to\infty}\frac{\log P(G_n({\bf X})>x)}{x^2}=-\frac{1}{2\max\limits_{1\le i\le n!}a^2_{ii}},$$
where $a^2_{ii}, i=1,2,\cdots, n! $ (factorial of $n$) are diagonal elements of matrix ${\bf C}{\bf \Sigma}{\bf C}'$ in Proposition 5.1 with $C_i=4i-2n-2, i=1,2,\cdots, n$.
\end{theorem}
{\bf Proof}\; By  Proposition 5.1,  if ${\bf{X}}\sim N_n({\boldsymbol \mu},{\bf \Sigma})$, then
${\bf{Y}}=(Y_1,\cdots,Y_m)' \sim N_m ({\bf C}{\boldsymbol \mu}, {\bf C}{\bf \Sigma}{\bf C}')$. In particular,
$Y_i\sim N_1 (\nu_i, a^2_{ii}), i=1,2,\cdots, m,$ where $\nu_i$'s are the elements of vector ${\bf C}{\boldsymbol \mu}$ and $a^2_{ii}$'s are diagonal elements of matrix ${\bf C}{\bf \Sigma}{\bf C}'$. Using the well known fact
 $$\int_x^{\infty}e^{-\frac{(z-\mu)^2}{2\sigma^2}}dz\thicksim\frac{\sigma^2}{x}e^{-\frac{x^2}{2\sigma^2}},\; x\to\infty,$$
we obtain, for all $1\le i\le n!$,
$$P(Y_i>x)\thicksim \frac{a_{ii}}{x}e^{-\frac{x^2}{2a^2_{ii}}},\; x\to\infty,$$
or, equivalently,
\begin{eqnarray*}
\lim_{x\to\infty}\frac{\log P(Y_i>x)}{x^2}=-\frac{1}{2a^2_{ii}}.
\end{eqnarray*}
Using (5.2) we get
\begin{eqnarray}
P(G_n({\bf X})>x)&=&\sum_{i=1}^m P(Y_j>x)-\sum\limits_{1\le i<j\le m}P(Y_i>x, Y_j>x)\nonumber\\
&&+\sum\limits_{1\le i<j<k\le m}P(Y_i>x, Y_j>x, Y_k>x)-\cdots\nonumber\\
&&+(-1)^m P(\cap_{i=1}^m \{Y_i>x\}).
\end{eqnarray}
Without loss of generality we assume that $a^2_{11}>a^2_{ii}, i=2,\cdots, m$. Then (5.4) can be rewritten as
 \begin{eqnarray*}
 P(G_n({\bf X})>x)=P(Y_1>x)(1+h(x)),
 \end{eqnarray*}
 where
 \begin{eqnarray*}
 h(x)&=&
  \sum_{i=2}^m \frac{P(Y_j>x)}{P(Y_1>x)}-\sum\limits_{1\le i<j\le m}\frac{P(Y_i>x, Y_j>x)} {P(Y_1>x)}\\
&&+\sum\limits_{1\le i<j<k\le m}\frac{P(Y_i>x, Y_j>x, Y_k>x)}{P(Y_1>x)}-\cdots\\
&&+(-1)^m \frac{P(\cap_{i=1}^m \{Y_i>x\})}{P(Y_1>x)}.
\end{eqnarray*}
One easily obtains
\begin{eqnarray*}
\lim_{x\to\infty}\frac{\log (1+h(x))}{x^2}=0.
\end{eqnarray*}
It follows from (5.3), (5.5) and (5.6) that
$$\lim_{x\to\infty}\frac{\log P(G_n({\bf X})>x)}{x^2}= -\frac{1}{2a^2_{11}},$$
as desired.

\section{ Concluding remarks}
In this paper, we have considered usual stochastic order and increasing convex order problems about  the Gini  mean differences for  multivariate elliptical random variables. The related issues for multivariate normal risks and scale mixture of multivariate normal risks   have been studied by  Samanthi et al. (2016) and Kim and Kim (2019). Here,  we have investigated  issues  for  multivariate elliptical risks and scale mixture of multivariate elliptical risks. This paper also  answers  the following open problems proposed in the Concluding Remarks in   Samanthi et al. (2016):  To what extent can Gini  mean differences of multivariate elliptical risks be ordered in the sense of usual stochastic order? Does
the conclusion still hold for high dimensional risks with general
elliptical distribution?   We  also solve  another  open problem  in Samanthi et al. (2016)  about  the increasing convex order of Gini  mean differences for higher dimensional risks.  In addition, we find the    tail
probability of Gini  mean difference $G_n({\bf X})$ when  ${\bf{X}}\sim {N}_n({\boldsymbol \mu},{\bf \Sigma})$. Especially,  a large deviation result for  Gini  mean difference of multivariate
normal risks is established which revised the corresponding result in Kim and Kim (2019).

\noindent{\bf\Large Acknowledgements}
  The research was supported by the National Natural Science Foundation of China (Nos. 12401616, 12301605).

\noindent {\bf Conflict of Interest} The authors declare that they have no known competing financial interests or personal relationships that could have appeared to influence the work reported in this paper.

\end{document}